\documentstyle[12pt]{article}
\input{psfig}
\begin{document}

\title{Leading particle effect, inelasticity and the connection 
between average multiplicities in {\bf $e^+e^-$} and {\bf $pp$}
processes}

\author{{M. Batista}\\
{\small Instituto de F\'{\i}sica
Te\'orica {\em IFT}}\\ {\small Universidade Estadual Paulista, Unesp}\\
{\small 01405-000 \  S\~ao Paulo \  SP \  Brazil \bigskip}
\\ and \vspace{0.25cm} \\
{R. J. M. Covolan} \\
{\small Instituto de F\'{\i}sica {\em Gleb Wataghin}} \\ {\small Universidade
Estadual de Campinas, Unicamp} \\ {\small 13083-970 \  Campinas \  SP \
Brazil \bigskip}}

\maketitle

\begin{abstract}

The Regge-Mueller formalism is used to describe the inclusive 
spectrum of the proton in $p p$ collisions. From such a 
description the energy dependences of both  average 
inelasticity and leading proton multiplicity are calculated. 
These quantities are then used to establish the connection
between the average charged particle multiplicities measured 
in {\bf $e^+e^-$} and {\bf $pp/{\bar p}p$} processes. The 
description obtained for the leading proton cross section 
implies that Feynman scaling is strongly violated only at 
the extreme values of $x_F$, that is at the central region 
($x_F \approx 0$) and at the diffraction region 
($x_F \approx 1$), while it is approximately observed 
in the intermediate region of the spectrum.

\end{abstract}

\newpage

\section{Introduction}

It is experimentally well known that the energy dependence 
of the charged particle multiplicities in $e^+ e^-$ and 
$pp/{\bar p}p$ processes exhibit a quite similar behavior. 
In the late 70's, experiments analysing $p p$ collisions 
at the CERN ISR Collider \cite{basile} have shown that 
not the total center-of-mass energy $\sqrt s$ is used 
for particle production; instead, a considerable fraction 
of the available energy is carried away by the leading proton. 
These experiments have shown that a more adequate way of 
comparing average multiplicities from different reactions 
is in terms of the amount of energy effectively used for 
multiparticle production. The problem is how to determine 
this quantity.

Observations like these have inspired several attempts 
to describe ${\langle n_{ch} \rangle}_{e^+e^-}$ and 
${\langle n_{ch} \rangle}_{pp}$ by an universal function. 
In ref.\cite{pol}, for instance, 
two corrections are made to compare these quantities: 
the energy variable for ${\langle n_{ch} \rangle}_{pp}$ 
is corrected by removing the portion referring to the 
elasticity (the fraction of the energy taken by the 
leading particle) and then the average leading proton 
multiplicity is subtracted. A similar idea is followed 
in ref.\cite{ref1} where attempts are made to establish 
this universal behavior
by fitting. 

In the present paper, we analyze the same subject by 
following an analagous point of view, but rephrasing 
the procedure in the following way. It is assumed that, 
if in $e^+ e^-$ collisions the average charged particle 
multiplicity is given by

\begin{equation}
{\langle n_{ch} \rangle}_{e^+e^-}=N(\sqrt{s}),
\label{mult1}
\end{equation}
then in ${pp}/{\bar p}p$ collisons we have
\begin{equation}
{\langle n_{ch} \rangle}_{pp}=\langle n_0 \rangle + 
N(\langle k_p\rangle\sqrt{s}),
\label{mult2}
\end{equation}
where $N(W)$ is an universal function of the energy 
available for multiparticle production, 
$W$, $\langle n_0 \rangle$ is the average leading 
particle multiplicity, and $\langle k_p\rangle$ is 
the average inelasticity.
In \cite{pol} and \cite{ref1}, the quantities related 
to $\langle n_0 \rangle$ and $\langle k_p\rangle$ are 
supposed to be constant. In particular, in \cite{ref1} 
they  are determined by a simultaneous fit of ${\langle 
n_{ch} \rangle}_{e^+e^-}$ and 
${\langle n_{ch} \rangle}_{pp}$ data. 

Our procedure, instead, consists in obtainning these 
quantities ($\langle n_0 \rangle$ and $\langle k_p\rangle$) 
not from fitting ${\langle n_{ch} \rangle}_{e^+e^-}$ 
and ${\langle n_{ch} \rangle}_{pp}$, but in a totally 
independent way, from the inclusive reaction 
$p p \rightarrow p X$, paying particular attention 
to their energy dependence. After doing that, the 
obtained $\langle n_0 \rangle$ and $\langle k_p\rangle$ 
are applied to (\ref{mult2}) via a parametrization of 
(\ref{mult1}) 
and the result is compared to data in order to verify 
to what extent such a hypothesis is acceptable.

This procedure seems to be very well defined and 
straightforward, but it should be noticed that it 
drives to some difficult problems. The question is 
that it requires a previous knowledge about the 
energy dependence of the inelasticity and about 
the behavior of the average leading particle 
multiplicity which  constitute  themselves problematic 
subjects.
In particular, the energy dependence of the average 
inelasticity is a very much  disputed question since 
there are opposite claims that this quantity increases \cite{nos,inc,nik,gaisser} or  that it decreases 
\cite{igm,dec,he} with increasing energy  at quite 
different rates. In spite of the of models predicting 
extreme behaviors, {\it i.e.} very fast increase of 
the inelasticity (like in \cite{nik}) or very fast 
decrease (like in \cite{igm}), most of these analyses 
referred here report the average inelasticity as 
having a smooth and slowly changing behavior.
\footnote{For a recent account on this subject from 
the viewpoint of cosmic-ray data, see  
ref.\cite{bellandi_novo}.} This is once again verified 
here in a different way.

The idea of discussing the energy behavior of the 
average inelasticity in connection with the energy 
dependence of $\langle n_0 \rangle$ and 
$\langle k_p\rangle$ is not new. Of particular 
interest to present work is an analysis with this 
purpose performed by He \cite{he}. He has extracted 
values of the average inelasticity by using arguments 
similar to those given above and obtained results 
pretty much in agreement with the predictions of 
ref.\cite{igm}. We shall argue below that such an 
agreement is probably due to the fact that two 
important effects are missing in his analysis.

Another controversial question involved in the present 
analysis (but treated here just {\it en passant}) is 
that referred to unitarity violation in diffractive 
dissociation processes. This is an old-standing problem  
that has come back to the scene due to the fact that 
recent measurements on hard diffractive production of 
jets and W's revealed a large discrepancy between 
data and  theoretical predictions.
In ref. \cite{dino}, it is proposed that such a discrepancy 
in hard diffraction has to do with unitarity violation in 
single diffractive processes. Since we are going to describe  
the inclusive reaction $p p \rightarrow p X$, we have to 
face this problem in the region of the spectrum where 
diffractive processes are dominant. 

A by-product of the present analysis is a complete 
parametrization for the reaction $p p \rightarrow p X$ 
in the whole phase space. This is obtained basically 
within the Regge-Mueller approach \cite{collins},  
but including the modifications suggested in \cite{dino} 
for the diffractive contribution.

The paper is organized as follows. In Section 2 we 
present the theoretical framework used to describe the 
leading particle sprectrum. Section 3 is devoted to 
show how this formalism is applied to describe the 
experimental data. In Section 4 we discuss the 
connection between ${\langle n_{ch} \rangle}_{e^+e^-}$ 
and ${\langle n_{ch} \rangle}_{pp}$. Our main 
conclusions are summarized in Section 5.

\section{\bf Leading particle spectrum}

In order to perform our analysis, we need to 
calculate the quantities
\begin{equation}
\langle n_0 \rangle=\frac{1}{\sigma_{inel}} 
\int{\frac{d\sigma}
{dx_F}\ dx_F}
\label{n0}
\end{equation}
and
\begin{equation}
\langle k_p \rangle  =  1 - \langle x_F 
\rangle 
= 1-\frac{1}{\sigma_{incl}}\ \int{x_F} \ 
\frac{d\sigma} {dx_F}\ dx_F
\label{kp}
\end{equation}
as a function of energy. We apply the Landshoff 
parametrization \cite{mul4}   
$\sigma_{inel}(s) = 56\ s^{-0.56} + 18.16\ s^{0.08}\ [mb]$ 
to represent  the inelastic cross section within the energy
 range where multiplicity data are included, and the 
inclusive cross section is simply given by 
$\sigma_{incl}\equiv \int{\frac{d\sigma} {dx_F}}\ dx_F$. 
Thus, the whole analysis depend on the knowledge of 
the leading particle spectrum ${d\sigma}/{dx_F}$ 
and its evolution with energy. The obtainment of 
this spectrum is detailed in the discussion that follows.

The invariant cross section for the inclusive reaction 
$a b \rightarrow c X$ is given by  
\begin{equation}
E\frac{d^3\sigma}{d{\bf p}^3} = \frac{2 E}{\pi 
{\sqrt s}}\frac{d^{3}\sigma} {dx_F\ dp_{T}^{2}}
\end{equation}
where $x_F = 2 p_L /{\sqrt s}$ is the Feynman variable 
for the produced particle $c$ and $E,\ p_L,\ p_T$ are 
respectively its energy, longitudinal and tranversal
 momenta. Particularly in the diffractive region 
($x_F \approx 1$) such a quantity is usually 
expressed in terms of
\begin{equation}
E\frac{d^3\sigma}{d{\bf p}^3} = \frac{s}{\pi} 
\frac{d^2\sigma}{dt\ dM^2} = 
 \frac{x_E}{\pi x_F} \frac{d^2\sigma}{dt\ d\xi},
\end{equation}
with  $x_E = 2 E /{\sqrt s}$, $\xi = M^2/s = 1-x_F$ 
and $-t = m_c^2\ {(1-x_F)^2}/{x_F} + {p_T^2}/{x_F}$. 
Variable $M^2$ is the missing mass squared  defined 
as $M^2 \equiv (p_a + p_b - p_c)^2$.

The procedure to calculate  the invariant cross section 
employed here comes from  the Regge-Mueller formalism 
which consists basically of the application of the 
Regge theory for hadron interactions to the 
Mueller's generalized optical theorem. This theorem 
establishes that the inclusive reaction 
$a b \rightarrow c X$ is connected to the elastic 
three-body amplitude $A (a b {\bar c} \rightarrow a 
b {\bar c})$ via 
\begin{equation}
E\frac{d^3\sigma}{d{\bf p}^3} (a b \rightarrow c X) 
\sim \frac{1}{s} Disc_{M^2} \ A (a b {\bar c} 
\rightarrow a b {\bar c}),
\label{disc}
\end{equation}
where the discontinuity is taken across the $M^2$ cut 
of the elastic amplitude.
It is assumed that this amplitude in turn is given by
the Regge pole approach. Different kinematical limits
imply in specific formulations for the invariant 
cross section at the fragmentation and central regions. 
In the following, we specify the concrete expressions 
that these formulations assume in such regions 
(details can be found in \cite{collins}).

\centerline{\bf A. Fragmentation Region}

In our description, the invariant cross section for
the reaction $p p \rightarrow p X$ at the fragmentation
region is compounded of three predominant contributions 
which are determined within the Triple Reggeon Model 
(this is the particular formulation that (\ref{disc}) 
assumes in the beam fragmentation region with the 
limits $M^2 \rightarrow \infty$ and  $s/M^2 \rightarrow 
\infty$ \cite{collins}). These contributions, 
depicted in Fig.2, correspond to pomeron, pion and 
reggeon exchanges and are referred to as $\rm I\!P 
\rm I\!P \rm I\!P$, $\pi \pi \rm I\!P$, $\rm I\!R 
\rm I\!R \rm I\!P$, respectively. 

In the diffractive region, the $\rm I\!P \rm I\!P 
\rm I\!P$ contribution is dominant and (we assume 
for the reasons given below) is given by

\begin{equation}
\left (\frac{d^2\sigma}{dtd\xi}\right )_{\rm I\!P \rm 
I\!P \rm I\!P}
=f_{\rm I\!P, Ren}(\xi,t)\times \sigma_{\rm I\!P p} 
(s\xi)
\label{mult6}
\end{equation}
where $f_{\rm I\!P, Ren}(\xi,t)$ is  the 
{\it renormalized} pomeron flux factor proposed 
in \cite{dino} with the parameters defined in 
\cite{covolan}, that is 

\begin{equation}
f_{\rm I\!P, Ren}(\xi,t) = \frac{f_{\rm I\!P}
(\xi,t)}{N(s)}
\label{renf}
\end{equation}
with the Donnachie-Landshoff flux factor \cite{dl}
\begin{equation}
f_{\rm I\!P}(\xi,t)=\frac{{\beta}_{0}^{2}}{16\pi}
F_{1}^2(t)\ \xi^{1-2{\alpha}_{\rm I\!P}(t)}
\label{dlf}
\end{equation}
and 
\begin{equation}
N(s)=\int_{1.5/s}^{1} \int^{t=0}_{-\infty} 
f_{\rm I\!P}(\xi,t)\  dt\ d\xi.
\end{equation}
In the above expressions, $F_1(t)$ is the Dirac 
form factor, 
\begin{equation}
F_1(t) = \frac{(4m^2-2.79t)}{(4m^2-t)}\ \frac{1}
{(1-\frac{t}{0.71})^2},
\end{equation}
the pomeron trajectory is $\alpha_{\rm I\!P}(t)=
1+\epsilon +\alpha^{'}t$ with  
 $\epsilon=0.104$, $\alpha^{'}=0.25\ GeV^{-2}$ and 
$\beta_0=6.56\ GeV^{-1}$, determined from \cite{covolan2}.
In Eq.(\ref{mult6}), the pomeron-proton cross section 
is given by 
\begin{equation}
\sigma_{\rm I\!P p} (M^2) = \beta_{0}\ g_{\rm I\!P}\ 
(s\xi )^{\epsilon}
\end{equation}
with the triple pomeron coupling determined from 
data as $g_{\rm I\!P}=1.21\ GeV^{-1}$.

Since this scheme to calculate the diffractive 
contribution is not the usual one, some comments are 
in order. The usual derivation of the Triple Pomeron 
Model gives (\ref{dlf}), the {\it standard flux factor}, 
instead of (\ref{renf}), the renormalized one. 
The problem is that the standard flux factor drives 
to strong unitarity violation and the {\it renormalization} 
procedure was conceived \cite{dino} as an {\it ad hoc} way to 
overcome this difficulty. Although a rigorous 
demonstration of the renormalized scheme is still 
missing, it is acceptable in the sense that it 
provides a good description for the experimental 
data at the diffractive region (see a detailed 
discussion in \cite{dino_novo}). 

The pion contribution ($\pi \pi \rm I\!P$) is given 
by \cite{field}

\begin{equation}
\left (\frac{d^2\sigma}{dtd\xi}\right )_{\pi \pi 
\rm I\!P } = 
f_{\pi}(\xi,t) \times \sigma_{\pi p}(s\xi)
\label{mult7}
\end{equation}
where
\begin{equation}
f_{\pi}(\xi,t) = \frac{1}{4\pi}\frac{g^2}{4\pi}\
\frac{|t|}{(t-\rho^2)^2}\ e^{b_{\pi} (t-\rho^2)}
\xi^{1-2{\alpha}_{\pi}(t)}
\end{equation}
and $\alpha_{\pi}(t)=0.9(t-\rho^2)$ with 
$\rho^{2}=m_{\pi}^{2}=0.02\ GeV^2$.
We follow \cite{robinson} in fixing the coupling 
constant in $g^2/4\pi=15.0$ and putting $b_{\pi}=0$ 
(see also \cite{field}). The pion-proton cross section 
$\sigma_{\pi p}(s\xi)=10.83\ (s\xi)^{0.104}\ +\ 27.13\ 
(s\xi)^{-0.32}\ [mb]$ is  taken from \cite{covolan2}.

If one considers only the diffractive and 
near-to-diffractive regions and low $p_T$ 
($-t \sim 0.0 - 0.1\ GeV^2$), the contributions 
outlined above are enough to provide a good 
description of the available data (see \cite{dino_novo}). 
However, when one wants to consider larger $p_T$ 
and $x_F < 0.9$, at least a third contribution is 
required. That is the reason why we introduce the reggeon 
contribution.

The reggeon contribution ($\rm I\!R \rm I\!R \rm I\!P$) 
is determined  by

\begin{equation}
\left (\frac{d^2\sigma}{dtd\xi}\right )_{\rm I\!R \rm 
I\!R \rm I\!P} = 
f_{\rm I\!R}(\xi,t) \times \sigma_{\rm I\!R p}(s\xi)
\end{equation}
with 
\begin{equation}
f_{\rm I\!R}(\xi,t) = \frac{{\beta}_{0{\rm I\!R}}^{2} }
{16\pi}
e^{2b_{\rm I\!R}  t}\ \xi^{1-2{\alpha}_{{\rm I\!R}}(t)},
\label{mult8}
\end{equation}
and
\begin{equation}
\sigma_{\rm I\!R p}(s\xi) = {\beta}_{0{\rm I\!R}}\ 
g_{{\rm I\!R} }(s\xi )^{\epsilon}.
\label{sigreg}
\end{equation}
In this case, the trajectory is assumed to be 
$\alpha_{{\rm I\!R}}(t)=0.5+t$ while  the  
constants $\beta_{{\rm I\!R}}\equiv 
({\beta}_{0{\rm I\!R}  }^{3}\ g_{\rm I\!R})$
and $b_{{\rm I\!R}}$ remain to be determined 
from data. 

Thus, with the expressions and parameters given 
above, the $\rm I\!P \rm I\!P \rm I\!P$ and  
$\pi \pi \rm I\!P$ contributions are completely 
specified; only the $\rm I\!R \rm I\!R \rm I\!P$ 
contribution remains to have the final parameters 
determined.

\centerline{\bf B. Central Region}

In order to describe the leading particle spectrum 
in the central region, we use the Double Reggeon Model 
\cite{collins} that gives the invariant cross section as

\begin{equation}
E\frac{d^3\sigma}{d{\bf p}^3}=\sum_{i,j}\ 
\gamma_{ij}(m_{T}^{2})
\ \left|\frac{t}{s_0}\right |^{\alpha_i(0)-1}\ \left |
\frac{u}{s_0}\right |^{\alpha_j(0)-1}
\label{mult9}
\end{equation}
where $m_{T}=({p_{T}^{2}+m^2_p})^{1/2}$ is the transversal 
mass,  and 
$u=-m_T\sqrt{s}\ e^{-y}$ and  $t=-m_T\sqrt{s}\ e^y$ are 
the Mandelstam variables given in terms of the rapidity 
$y=ln \frac{(E+p_L)}{m_T}$. Function  
$\gamma_{ij}(m_{T}^{2})$
corresponds to the product of the three vertices of 
the diagrams depicted in Fig.3. These diagrams 
represent the contributions taken into account in the 
present analysis: $\rm I\!P \rm I\!P$, $\rm I\!P 
\rm I\!R + \rm I\!R 
\rm I\!P$, and $\rm I\!R \rm I\!R$ (pion contributions 
are not considered in this case because they are 
totally covered by the others).

We assume for the coupling function  $\gamma_{ij}
(m_{T}^{2})$ a simple gaussian form, 
\begin{equation}
\gamma_{i j}(m_{T}^{2})=\Gamma_{i j}
\ e^{-a_{i j}m_{T}^{2}}
\end{equation}
where $\Gamma_{i j}$ is a constant that already 
embodies the product of the couplings belonging 
to the triple and quartic vertices. With these 
definitions, the invariant cross sections for 
the three contributions become 

\begin{equation}
\left (E\frac{d^3\sigma}{d{\bf p}^3} \right )_
{\rm I\!P \rm I\!P} = \Gamma_{\rm I\!P \rm I\!P}
\ e^{-a_{\rm I\!P \rm I\!P}m_{T}^{2}}\ 
(m_T\ \sqrt{s})^{2\epsilon},
\label{mult10}
\end{equation}

\begin{eqnarray}
\left (E\frac{d^3\sigma}{d{\bf p}^3} \right )_
{\rm I\!P {\rm I\!R}  +{\rm I\!R}   \rm I\!P}=2\ 
\Gamma_{\rm I\!P {\rm I\!R}}\ e^{-a_{\rm I\!P 
{\rm I\!R}  }m_{T}^{2}} \ (m_T\ \sqrt{s})^
{\epsilon+\alpha_{{\rm I\!R}  }(0)-1}
\  cosh[(1+\epsilon-\alpha_{\rm I\!R}  (0))y], 
\label{mult11}
\end{eqnarray}
and

\begin{equation}
\left (E\frac{d^3\sigma}{d{\bf p}^3} \right )_
{{\rm I\!R}  {\rm I\!R}  }= \Gamma_{{\rm I\!R}  
{\rm I\!R}}\ e^{-a_{{\rm I\!R}  {\rm I\!R}  }
m_{T}^{2}}\ (m_T\ \sqrt{s})^{2(\alpha_
{\rm I\!R}  (0)-1)}.
\label{mult12}
\end{equation}
In the above expressions again $\alpha_{\rm I\!R}(0)=0.5$ 
and $\epsilon=0.104$ \cite{covolan2}. Differently 
from the fragmentation region where almost all 
parameters are already established, in this 
region almost all of them (expect for the 
intercepts just mentioned) must be fixed 
from data.

The expressions given above could be enriched
by detailing the reggeon exchange in terms 
of $f$, $\rho$, $\omega$, $a_2$, and taking 
into account all crossed terms, but in fact 
we are pursuing here a minimal description 
in which only the dominant and effective 
contributions are considered. We shall 
see below that these contributions are 
enough to provide a good description of 
the available data.

\section{\bf Description of experimental data}

Experimental data on leading particle spectrum 
are very scarce. A compilation for 
$p p \rightarrow p X$ is shown in Fig.1  
where data from three experiments 
\cite{basile,aguilar,brenner} are put together 
(the curve and the insert in this figure should 
be ignored for the moment). 
As can be seen, a pretty flat spectrum is exhibited, 
except for $x_F \approx 1$ where the typical 
diffractive peak appears.\footnote{This peak 
is absent from the Aguilar-Benitez 
{\it et al.} data due to trigger inefficiency 
for $x_F > 0.75$ in this particular 
experiment \cite{aguilar}.}

The problem that arises when one tries to describe 
the $p p \rightarrow p X$ reaction in the whole 
phase space is that the available data are not 
enough to determine unambigously each one of the 
contributions  outlined above. One may have noted 
in the previous section that we have summarized 
all secondary reggeon exchanges (except for the pion) 
in a single contribution denoted by $\rm I\!R$ and 
the reason is the following. When one analyzes, 
for instance, total cross section data (like in 
\cite{covolan2}), it is possible to establish 
(to a certain extent) the relative amount of 
the different contributions. Actually, this is 
enforced by the changing shape exhibited by the 
data in different regions. That is not the case 
here because out of the diffractive region the 
spectrum is pretty flat and that makes it difficult 
to discriminate the regions where the different 
exchange processes contribute the most. Thus, 
in order to establish how the expressions 
outlined above are summed up to compose the 
observed spectrum, we have to follow a 
particular strategy.

Since our intention was obtainning an acceptable 
description for $p p \rightarrow p X$ data in the 
whole phase space, we did not use in our fitting 
procedure the data shown in Fig.1 which represent 
only the $x_F$-dependence. Instead, we have set 
those data apart to be used only at the end to 
check our final results which, in fact, were 
obtained with distributions giving in terms 
of both $x_F$ and $p_T$ dependences.

Our procedures to determine the contributions 
at the central and at the fragmentation regions 
are quite different. The main problem is that 
these regions overlap each other and thus it 
is pratically impossible to separate them 
(or establish clear limits). To overcome this difficulty 
we assumed that, except  for normalization effects, 
the $x_F$ and $p_T$ dependences of the proton produced 
in the central region through the reaction  
$p p \rightarrow p X$ is the same as for the antiproton 
produced in  $p p \rightarrow {\bar p} X$. This 
assumption was implemented by fitting simultaneously 
the data shown in Figs. 4 and 5 \cite{dados1,dados2} 
through the expressions

\begin{eqnarray}
\left (E\frac{d^3\sigma}{d{\bf p}^3} \right )_{pp->
\bar{p}X}^{central}=
\left (E\frac{d^3\sigma}{d{\bf p}^3} \right )_
{\rm I\!P \rm I\!P}
+\left (E\frac{d^3\sigma}{d{\bf p}^3} \right )_
{\rm I\!P {\rm I\!R}  +{\rm I\!R}  \rm I\!P}
+\left (E\frac{d^3\sigma}{d{\bf p}^3} \right )_
{{\rm I\!R}  {\rm I\!R}  }
\label{mult12b}
\end{eqnarray}
and
\begin{equation}
\left (E\frac{d^3\sigma}{d{\bf p}^3} \right )_
{pp->pX}^{central}=\lambda(s)\
\left (E\frac{d^3\sigma}{d{\bf p}^3} \right )_
{pp->\bar{p}X}^{central}.
\label{mult13}
\end{equation}
The idea is that the data of Fig.4 provide the 
information on the $x_F$ and $p_T$ dependences 
through Eqs. (\ref{mult10})-(\ref{mult12b}) and 
relation
$x_F=2 m_T sinh(y)/{\sqrt s}$, while the connection 
between  $p p \rightarrow {\bar p} X$ and  
$p p \rightarrow p X$ is established by fitting 
the data of Fig.5 through the function 
$\lambda(s)$ of Eq.(\ref{mult13}). The parameters 
$\Gamma_{ij}$ and $a_{ij}$ of this fit are given 
in Table \ref{tabmul1}
while $\lambda(s)$ is parametrized as \  
$\lambda(s)=1.0+11.0\ s^{-0.3}$. 

The agreement with data of Figs.4 and 5 is not 
perfect, but that is because we are simplifying 
the description by considering only a few 
contibutions, the dominant ones. As stated 
before, this is enough for the purposes of the 
present analysis.

Now we are able to obtain the total description 
by adding up central and fragmentation region 
contributions. 
As explained before, the contributions dominant 
at the fragmentation region, 
Eqs. (\ref{mult6})-(\ref{sigreg}), are almost 
completely determined. The parameters 
$\beta_{{\rm I\!R}}$ and $b_{{\rm I\!R}}$ 
referring to the $\rm I\!R \rm I\!R \rm I\!P$ 
contribution are established by fitting the 
data of Fig.6 (from \cite{brenner}). This is 
done by using the expression 

\begin{eqnarray}
\left (E\frac{d^3\sigma}{d{\bf p}^3}\right )_
{pp->pX}^{total}=
\left (E\frac{d^3\sigma}{d{\bf p}^3}\right )_
{\rm I\!P \rm I\!P \rm I\!P}
+\left (E\frac{d^3\sigma}{d{\bf p}^3}\right )_
{\pi \pi \rm I\!P} \ 
+\left (E\frac{d^3\sigma}{d{\bf p}^3}\right )_
{\rm I\!R \rm I\!R \rm I\!P}
+\left (E\frac{d^3\sigma}{d{\bf p}^3} \right )_
{pp->pX}^{central}
\label{mult14}
\end{eqnarray}
where the last term refers to Eq.(\ref{mult13}) 
with the parameters given in Table \ref{tabmul1}. 
With this final fit the remaining parameters 
result to be 
$\beta_{{\rm I\!R}  }=2465.7\ mb\ GeV^{-2}$ and  
$b_{\rm I\!R}  =0.1\ GeV^{-2}$. 
Fig.7 offers a view of how the different contributions 
are composed to form the final result and how this 
picture evolves with $p_T$.

The different contributions of the invariant cross 
section in both regions integrated over $p_T$ 
produce the results of ${d\sigma}/{dx_F}$  for 
both reactions 
exhibited in Fig.1 (solid curves) for $p_{lab}=
400\ GeV/c$. We remind the reader that these 
data were not used in the fit, but are used 
now to check the reliability of the whole 
procedure.
From this figure it is possible to see that the 
final description obtained for the leading proton 
spectrum is quite reasonable.

\section{Connection between 
${\langle n_{ch} \rangle}_{e^+e^-}$ and 
${\langle n_{ch} \rangle}_{pp}$}

The results obtained above specify completely 
the behavior of the leading particle spectrum 
and allow us to calculate $\langle n_0 \rangle$ 
and $\langle k_p\rangle$  as given by (\ref{n0}) 
and (\ref{kp}).  In Fig.8, we show the energy 
dependence of these quantities as obtained in 
the present analysis (solid curves). In the 
same figure, it is also shown the average 
inelasticity as predicted by the Interacting 
Gluon Model (IGM) \cite{igm} (dot-dashed curve)
for comparison. The average inelasticity obtained 
from the present analysis is very slowly increasing 
with energy, close to the behavior predicted by the 
Minijet Model \cite{gaisser}.

With these results we can 
come back to our original intent which is checking 
the hypothesis of universal behavior of the multiplicity 
that is specified  by Eqs.(\ref{mult1}) and (\ref{mult2}). 
In order to do that, we first establish 
a parametrization for $N(\sqrt{s})$ through

\begin{equation}
N(\sqrt{s})=a_1+a_2\ ln(\frac{s}{s_{0}})+a_3\ 
ln^{2}(\frac{s}{s_{0}})
\label{mult15}
\end{equation}
with $s_{0}=1\ GeV^2$. However, before performing 
the fit to experimental data, an additional effect 
has to be considered. This is because, besides the 
charged particles produced at the primary vertex, 
${\langle n_{ch} \rangle}_{e^+e^-}$ data include 
also decay products of $K^0_s \rightarrow \pi^+ \pi^-$, 
$\Lambda \rightarrow p \pi^-$, and  ${\bar\Lambda} 
\rightarrow {\bar p} \pi^+$. Following \cite{ref1}, 
we take this contamination into account by computing 
the ratio $R={\langle n_{ch} \rangle}_{K^0_s,\Lambda,
{\bar\Lambda}}/{{\langle n_{ch} \rangle}_{e^+e^-}}$ 
and by redefining  (\ref{mult1}) as

\begin{equation}
{\langle n_{ch} \rangle}_{e^+e^-}=\frac{N(\sqrt{s})}{1-R},
\label{multc}
\end{equation}
with $R=0.097\pm 0.003$. This value was taken from 
\cite{ref1} and, besides the references quoted therein, 
it is in good agreement with experimental data from 
ref.\cite{check}. No energy dependence for $R$ 
can be inferred from these data.
The fit using (\ref{mult15}) and (\ref{multc}) 
gives $a_1 = 2.392$, $a_2=0.024$, and $a_3=0.193$.

In Fig.9,  we show the above parametrization 
describing $\langle n_{ch}\rangle _{e^+e^-}$ data 
from references quoted in \cite{ee}  and the 
calculated curve for $\langle n_{ch} \rangle _{pp}$ 
in comparison with data from \cite{pp}. The 
agreement with these data enables us to consider 
that our premises about the universal behavior of 
${\langle n_{ch} \rangle}_{e^+e^-}$ and 
${\langle n_{ch} \rangle}_{pp}$ are confirmed. 
Of course, this conclusion is restricted to the 
energy dependence of $\langle n_0 \rangle$ and 
$\langle k_p\rangle$ shown in Fig.8. 

The solid curve of the insert in Fig.9 shows what 
happens when the IGM average inelasticity is applied 
to the same purposes. One could argue that this last 
result is conditioned by the use of $\langle n_0 \rangle$ 
obtained in the present analysis which increases with energy. 
However, we note that increase in $\langle n_0 \rangle$ 
plays against increase in $\langle k_p \rangle$
since these are competitive effects.

Now a comment on the He analysis \cite{he}, where 
the relation
\begin{equation}
n_{ch}^{e^+ e^-} ({\sqrt {s_{e^+ e^-}}}) = 
n_{ch}^{pp} (k({\sqrt {s_{pp}}}){\sqrt {s_{pp}}})
\label{eq_he}
\end{equation}
is employed. After fitting $n_{ch}^{e^+ e^-}$ and 
$n_{ch}^{pp}$ independently, 
He imposes that relation (\ref{eq_he}) holds and 
extracts the inelasticity $k$ 
from this assumption. This is similar to what we 
have done, but 
we think that the result of decreasing inelasticity  
and the agreement with IGM obtained in such an 
analysis comes from the fact that neither the 
leading particle multiplicity ($n_0$) nor the 
effect of decay products ($R$) is considered 
and we see no reason for ignoring such effects.

A surprising outcome of the present analysis is 
shown in Fig.10 (a) where the normalized cross 
section $1/{\sigma_{incl}}\ d\sigma/dx_F$ is 
calculated up to the LHC energy. It is shown that, 
if the present description holds up to such high 
energies, Feynman scaling is approximately 
observed in the intermediate fragmentation region, 
$0.2 < x_F < 0.8$, but is violated in opposite 
ways at the central and  diffractive regions. 
Fig. 10 (b) shows the same results but in a 
scale that makes more evident the scaling 
violation at the central region. This result 
seems to say that the increase of production 
activity at the central region occurs at the 
expenses of a supression of the diffractive 
processes. However, this is just a speculative
observation that should be investigated more 
thoroughly.

\section{Conclusions}

We have presented in this paper a description of 
the inclusive reaction $p p \rightarrow p X$ in 
the whole phase space within the Regge-Mueller 
formalism, modified by the renormalization of the 
diffractive cross section.
The average multiplicity and the average inelasticity
were obtained from the leading proton spectrum and 
both of them resulted to be increasing functions of
energy, in agreement with \cite{nos,inc} and
particularly with \cite{gaisser}. The energy dependence 
of these quantities is such that allows one to 
accommodate very well the charged particle 
multiplicities ${\langle n_{ch} \rangle}_{e^+e^-}$ 
and ${\langle n_{ch} \rangle}_{pp}$ by an universal 
function once an appropriate relation is used.

An additional result is that the normalized leading 
proton spectrum approximately observe Feynman scaling 
for intermediate $x_F$, whereas such scaling is 
violated at the central and diffractive regions.

\section*{Acknowledgementes}

We are grateful to J. Montanha for valuable discussions 
and suggestions. 
We would like to thank also the Brazilian governmental 
agencies CNPq and FAPESP
for financial support.

\begin{table}[htbp]
\centering
\begin{tabular}{|c|c|c|}
\hline
$ij$ & $\Gamma_{ij}\ (mb\ GeV^{-2})$ & $a_{ij}\ 
(GeV^{-2})$ \\ \hline
$\rm I\!P \rm I\!P$ & $23.53$ & $3.90$ \\
$\rm I\!P {\rm I\!R}  $ & $-29.8$ & $3.45$ \\
${\rm I\!R}  {\rm I\!R}  $ & $13.75$ & $1.80$\\ 
\hline
\end{tabular}
\caption{\label{tabmul1} { Values of the parameters 
$\Gamma_{ij}$ and $a_{ij}$.}}
\end{table}

\begin{figure}[htbp]
\centerline{\psfig{figure=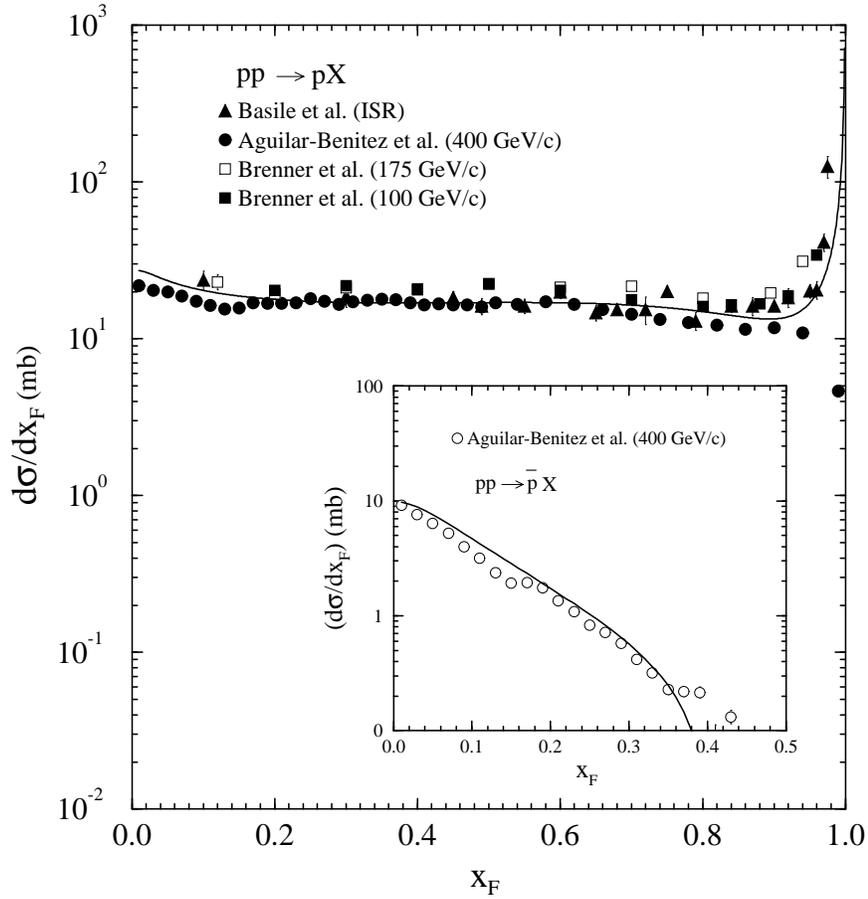,height=10.cm}}
\vspace{0.8cm}
\caption{Inclusive spectrum for the reactions 
$pp\rightarrow pX$ and $pp\rightarrow {\bar p}X$ 
(in the insert). Data from 
\protect{\cite{basile,aguilar,brenner}}. 
The solid curves are the results of the 
fit described in the text calculated for 
$400 \ GeV/c$.}
\end{figure}
\newpage
\begin{figure}[htbp]
\centerline{\psfig{figure=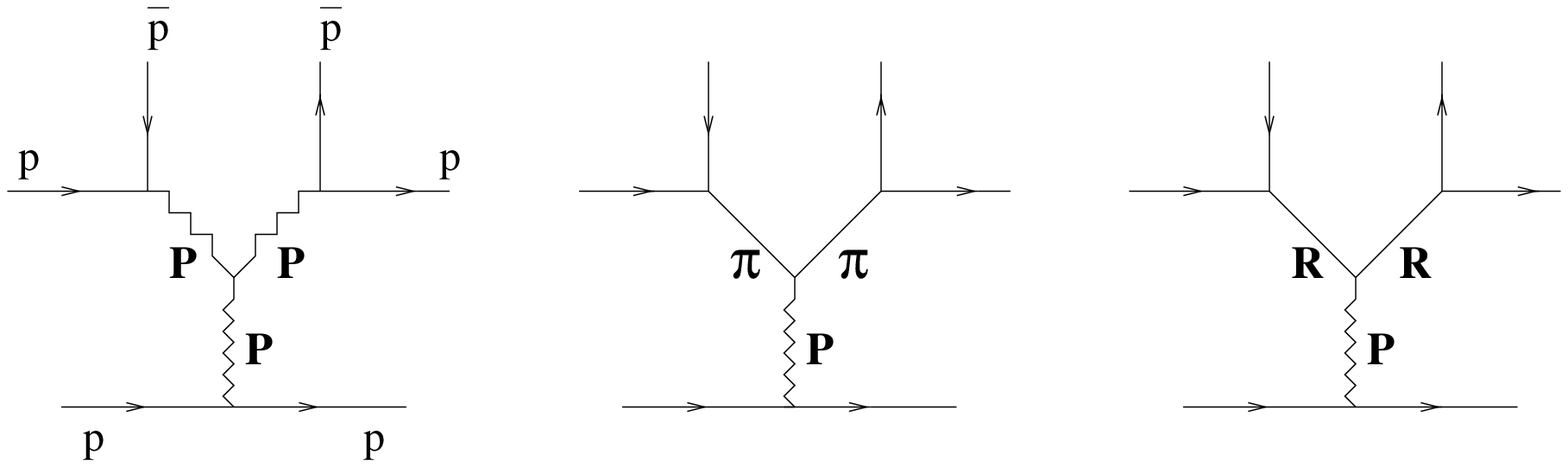,height=6.5cm}}
\vspace{0.8cm}
\caption{Triple-reggeon diagrams 
considered in the present analysis for the 
reaction $pp\rightarrow pX$. The particles 
corresponding to the external lines are all 
as in the first diagram. These diagrams
represent the contributions that are 
dominant in the fragmentation region.}
\end{figure} 

\begin{figure}[htbp]
\centerline{\psfig{figure=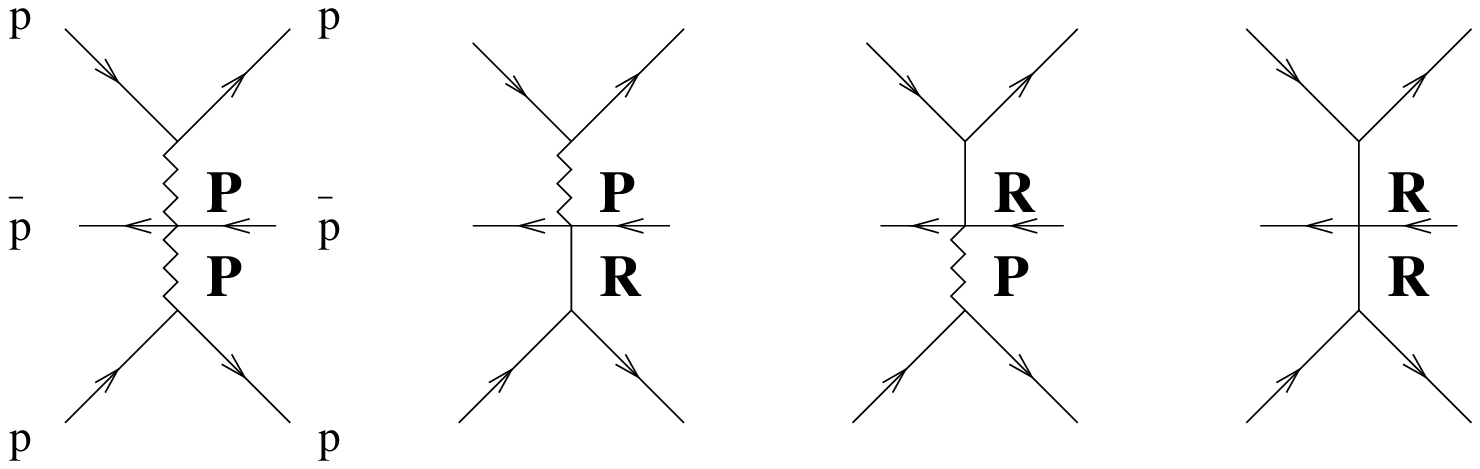,height=6.5cm}}
\vspace{0.8cm}
\caption{Double-reggeon diagrams 
considered in the present analysis for the 
reaction $pp\rightarrow pX$. The particles 
corresponding to the external lines are all 
as in the first diagram. These diagrams represent
the contributions  that are dominant in the central
region.}
\end{figure}

\begin{figure}[htbp]
\centerline{\psfig{figure=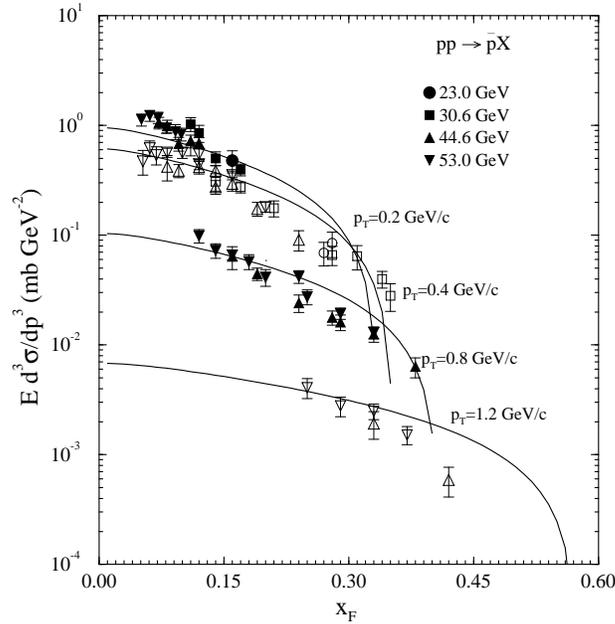,height=7.cm}}
\vspace{0.8cm}
\caption{Invariant cross section for 
the reaction {\protect $pp\rightarrow \bar{p}X$} at the 
ISR energies. The description is obtained with  Eqs.(\protect{\ref{mult10}})-(\protect{\ref{mult12b}}) and parameters 
of Table 1. Data taken from \protect \cite{dados1}.}
\end{figure}

\begin{figure}[htbp]
\centerline{\psfig{figure=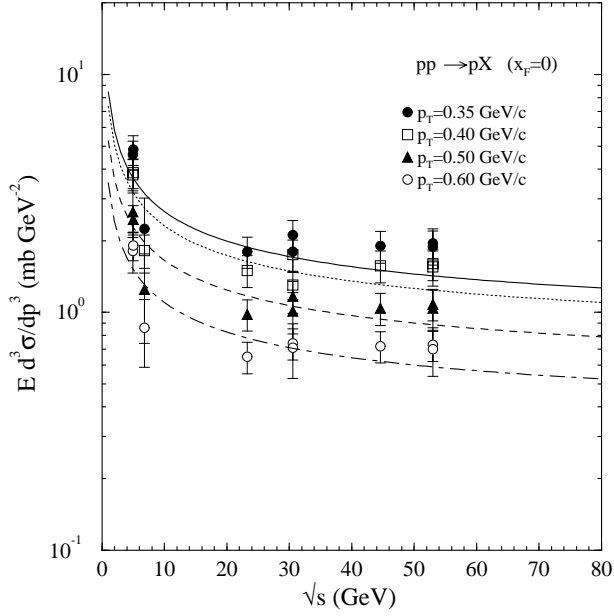,height=7.cm}}
\vspace{0.8cm}
\caption{Invariant cross section for 
the reaction $pp\rightarrow p X$ at $x_F =0$. 
The description is obtained with Eq.(\protect \ref{mult13}). 
Data taken from \protect \cite{dados2}.}
\end{figure}

\begin{figure}[htbp]
\centerline{\psfig{figure=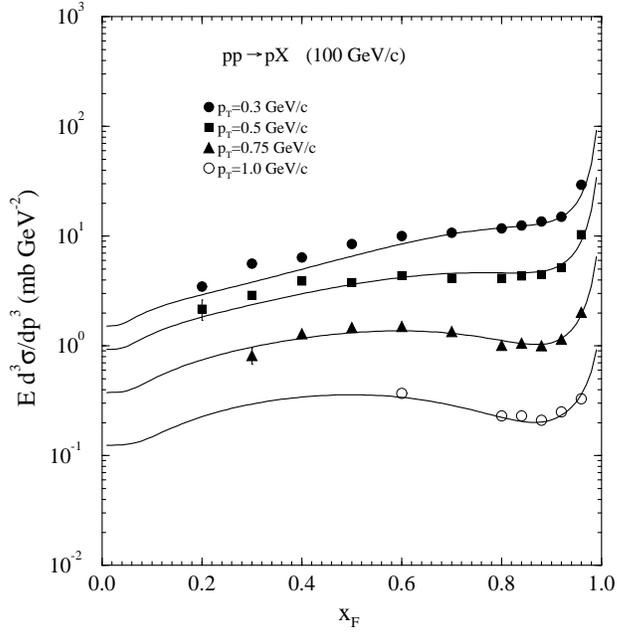,height=7.cm}}
\vspace{0.8cm}
\caption{Invariant cross section for the 
\protect{$pp\rightarrow p X$} at the fragmentation region. 
Curves calculated with Eq.(\protect \ref{mult14}).
Data from \protect \cite{brenner}.}
\end{figure}

\begin{figure}[htbp]
\centerline{\psfig{figure=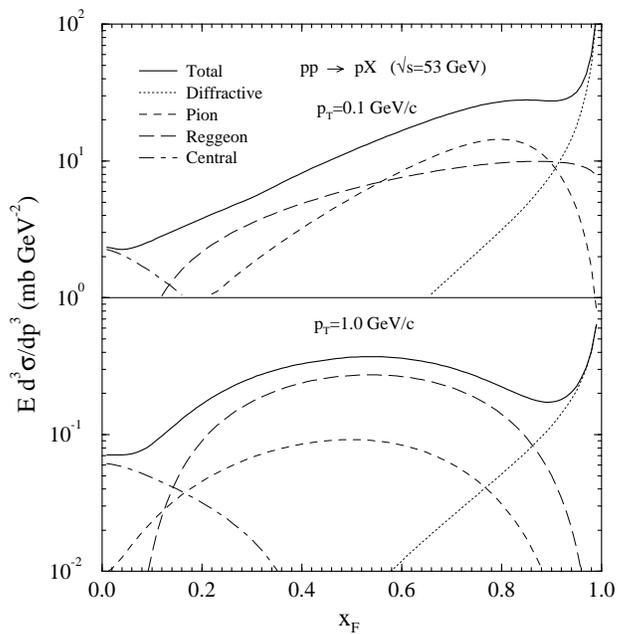,height=7.cm}}
\vspace{0.8cm}
\caption{Invariant cross section and 
its contributions for \protect $pp\rightarrow pX$ at 
$\protect \sqrt{s}= 53\ GeV$ for two \protect $p_T$  values.
These plots show how the 
interplay among the different contributions changes
as $p_T$ increases.}
\end{figure}

\begin{figure}[htbp]
\centerline{\psfig{figure=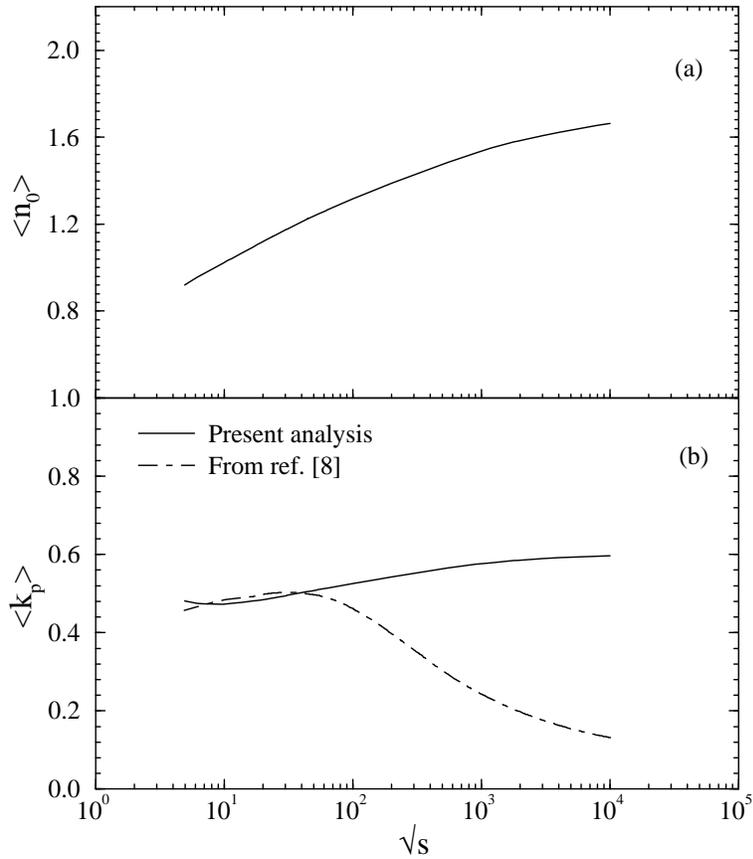,height=10.cm}}
\vspace{0.8cm}
\caption{Energy dependence of (a) average 
leading proton multiplicity
$\langle n_0 \rangle$ and (b) average inelasticity 
$\langle k_p \rangle$. In the lower figure (b)
it is shown $\langle k_p \rangle$  
obtained in the 
present analysis (solid curve) compared to the 
same quantity as predicted by the IGM \protect \cite{igm} 
(dot-dashed curve).}
\end{figure}

\begin{figure}[htbp]
\centerline{\psfig{figure=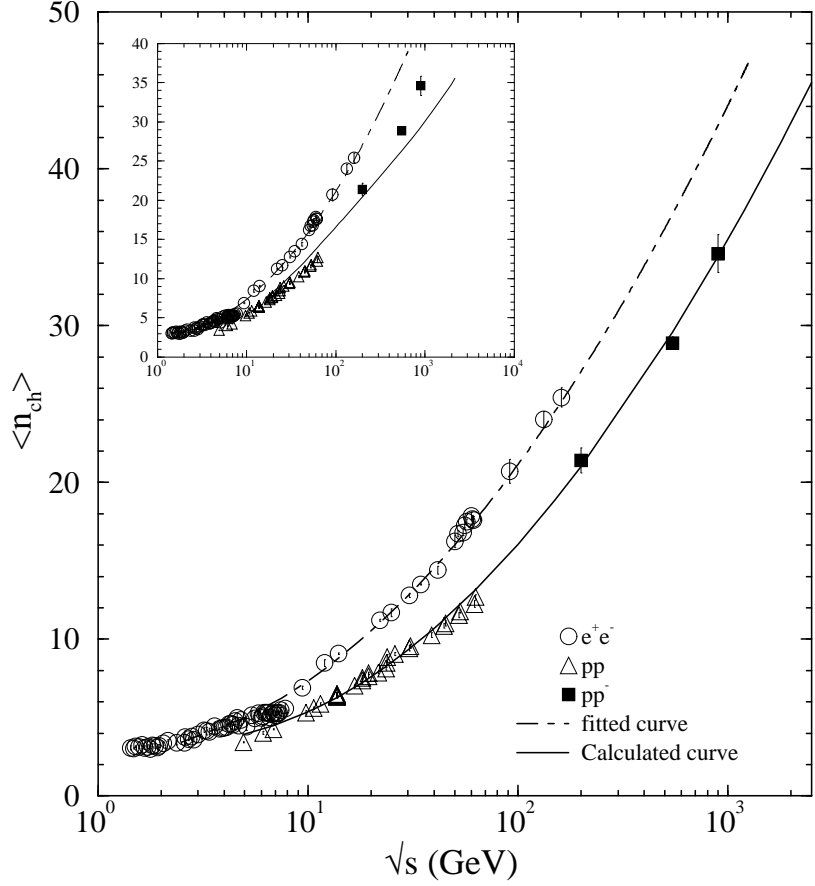,height=10.cm}}
\vspace{0.8cm}
\caption{Main figure: average charged 
particle multiplicities ${\langle n_{ch} \rangle}_{e^+e^-}$ and 
${\langle n_{ch} \rangle}_{pp}$ 
as a function of the center-of-mass energy. 
The dot-dashed curve refers to the fit obtained 
with Eqs.(\protect \ref{mult15}) and (\protect \ref{multc}). 
The solid curve was calculated with Eq.(\protect \ref{mult2}) 
by using  $\langle n_0 \rangle$ and 
$\langle k_p\rangle$ as calculated in the 
present analysis. Insert: the same as in the 
main figure, but using the average inelasticity 
given by the IGM \protect \cite{igm}.}
\end{figure}

\begin{figure}[htbp]
\centerline{\psfig{figure=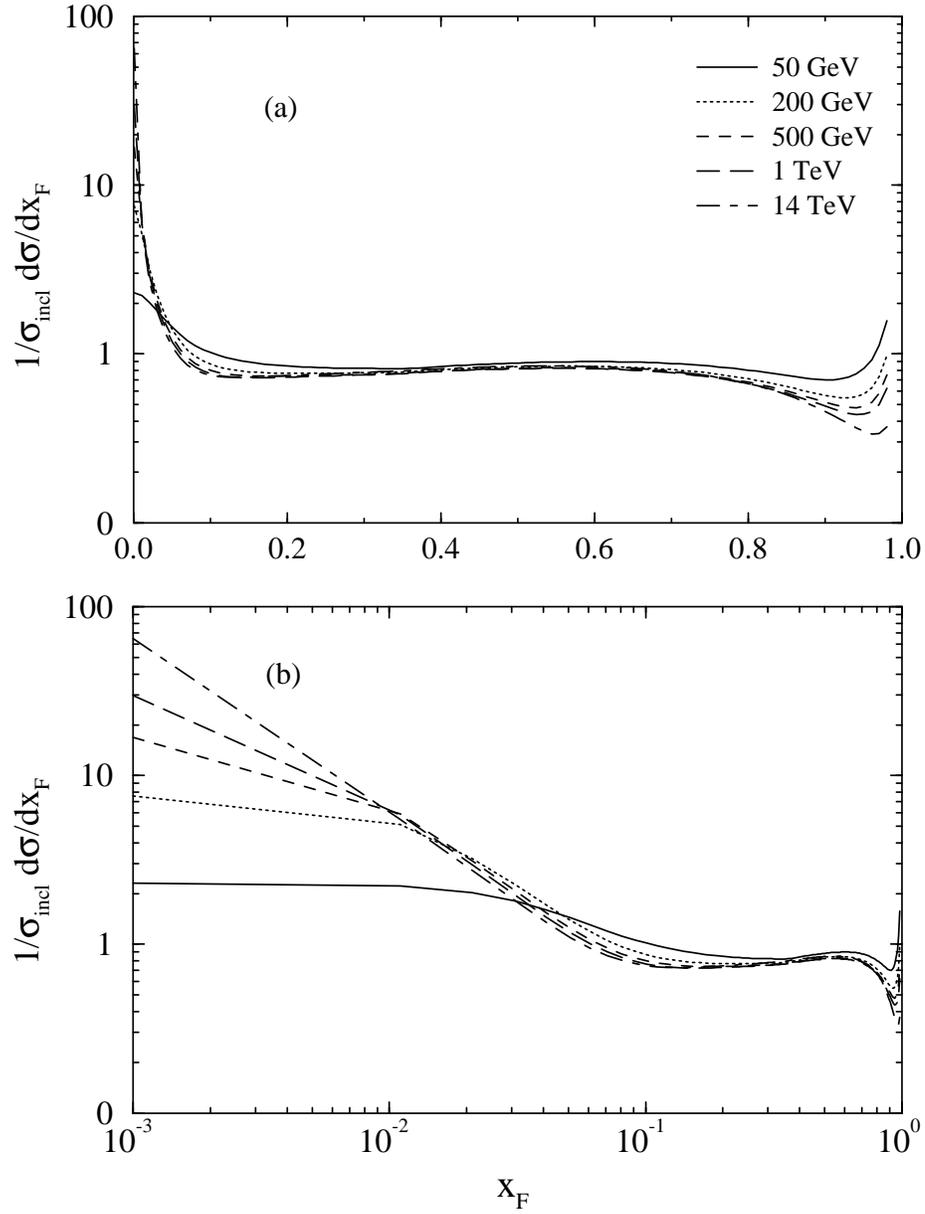,height=14.cm}}
\vspace{0.8cm}
\caption{(a) Normalized leading proton spectrum 
calculated up to the LHC energy. (b) The same  as (a)
but with logarithmic scale for $x_F$.}
\end{figure}

\end{document}